# Cloud Computing -An Approach with Modern Cryptography

Sugata Sanyal
**Corporate Technology Office**
**Tata Consultancy Services**
Mumbai, INDIA
sugata.sanyal@tcs.com

Parthasarathy P. Iyer (Corresponding Author)
**Self-Inventions-R&D**
Mumbai, INDIA
iyerparth@rediffmail.com

*Abstract*-In this paper we are proposing an algorithm which uses AES technique of 128/192/256 bit cipher key in encryption and decryption of data. AES provides high security as compared to other encryption techniques along with RSA. Cloud computing provides the customer with the requested services. It refers to applications and services that run on distributed network using virtualized resources and accessed by common IP and network standard. While providing data services it is becoming important to provide security for data. In cloud computing keeping data secure is an important issue to be focused. Even though AES was designed for military purposes, now a days it is been commercially adopted worldwide as it can encrypt most confidential document, as well as it can work in most restricted areas, and offers good defense against various attack techniques, and security level to protect data for next 2-3 decades.

*Keywords- Cryptography, AES, RSA, Cryptography for Cloud, CSP, Berkeley Solution*

I.   **Introduction** There are some general key mechanisms for protecting data: access control, auditing, authentication, authorization etc. In addition with this, cloud demands encryption process. The goal of encrypted storage in cloud is to create a virtual private storage system that maintains confidentiality and data integrity while maintaining the benefits of cloud storage. In the following sections we will see the AES method used for encryption [1].Cloud providers need to safeguard the privacy and security of personal data that they hold on behalf of users and organizations. In this paper we have assessed some of the key issues involved, and set out the basis of some approaches that we believe will be a step forward in addressing security issues [2].Cloud providers often have several powerful servers and resources in order to provide appropriate services for their users but cloud is at risk similar to other Internet-based technology. These issues which discussed in this paper are the main reasons that cause many enterprises which have a plane to migrate to cloud prefer using cloud for less sensitive data and store important data in their own local machines.

A.       **Tricks for Security** To advance cloud computing, the community must take proactive measures to ensure security. The Berkeley papers Solution is the data encryption. Before storing it at virtual location, encrypt the data with your own keys and make sure that a vendor is ready for security certifications and external audits [3] [4].
1. Identity management, access control, reporting of security incidents, personnel and physical layer management should be evaluated before you select a CSP (**Cloud Service Provider**). And you should minimize personal information sent to and stored in the cloud. CSP should maximize the user control and provide feedback
2. Organizations need to run applications and data transfer in their own private cloud and then transmute it into public cloud. While there are many legal issues exist in the cloud computing, Cloud Security Alliance should design relevant standards as quickly as possible.
3. Open Security Architecture (OSA) provides free frameworks that are easily integrated in applications, for the security architecture community. Its patterns are based on schematics that show the information traffic flow for particular implementation as well as policies implemented at each step for security reasons.
4. One thing to keep in mind are contract policies between clients and vendors, so that data belongs only to the client at all times, preventing third parties to be involved at any point. Also, authentication should be backed by several methods like password plus flash card, or password plus finger print, or some combination of external hardware and password.
5. For access Control, Security Assertion Markup Language (SAML) has been around for over 6 years now and is an excellent way of providing a Single Sign On solution across the enterprise firewall [5].

6. Examine your service level agreement. Ask your cloud computing vendor the tough questions: what roles and responsibilities will be theirs alone? How do they verify the safety of their systems? What immediate steps will be taken if there is a security breach?

7. Hire a third-party to complete an independent IT audit to identify your cloud computing weaknesses. An auditing firm should conduct interviews, observe and inspect processes and run penetration testing. The firm conducting the IT audit should have no stake one way or the other in the outcome.

## B.      Cryptography for Cloud

This standard specifies the Rijndael algorithm, a symmetric block cipher that can process data blocks of 128 bits, using cipher keys with lengths of 128, 192, and 256 bits. Rijndael was designed to handle additional block sizes and key lengths, and however they are not adopted in this standard.  Throughout the remainder of this standard, the algorithm specified herein will be referred to as "the AES algorithm." The algorithm may be used with the three different key lengths indicated above, and therefore these different "flavors" may be referred to as "AES-128", "AES-192", and "AES-256"[6][7]. AES algorithm uses a round function that is composed of four different byte-oriented transformations: 1) byte substitution using a substitution table (S-box), 2) shifting rows of the State array by different offsets, 3) mixing the data within each column of the State array, and 4) adding a Round Key to the State.

In AES the main architecture is centrally controlled by both hardware and software. Decryption of this system is depending upon the designing rule which is called Substitution Permutation Networking. Advanced Encryption standard has standard blocks with fixed length of 128 bits and their allowed key size is 128,192 or it can be 256 bits, new research has evolved that multiple key size can be allocated to the block it could be 32 bits with the least capacity of 128 bits and its key size may be extended no fixed length is announced. Its operations are based on the 4X4 matrix of the bytes with finite field calculations especially designed for the purpose of calculations. AES specifies the repetition numbers for converting the input to the normal readable text. An input provided by the user undergoes several steps of processing according to the encryption key provided. These steps are categorized as rounds and for converting the text back to the encrypted form same rounds are repeated with the same key which was used for decryption [8]. Numbers of repetitions are decided on the nature and level of the algorithm used as the base of system. Rounds are depending on the schedule of the algorithm provided by the AES key Page Style.

**II. Essence of Cryptography and Model**
**AES Process:**
Again, we start by looking at the overall structure of the AES cipher. In the case of AES, the block size is 128 bits and the key size can be 128, 192 or 256 bits. The original Rijndael specification also supported several block sizes, but in the AES standard itself only 128 bits blocks are defined.

Just like with DES, the cipher consists of a basic operation called *round* which is repeated a number of times. In this case, AES is based in a design principle called *Substitution-Permutation Networks* which means that the cipher is composed of a series of substitutions and permutations one after each other[9][10].

The number of *rounds (R)* in AES depends on the key length: 10 rounds for 128, 12 rounds for 192 and 14 rounds for 256 bits. AES works on a structure known as *the AES state*, which is simply an arrangement of the block in a 4x4 matrix. Furthermore, most AES operations can be described as operations in the $GF(2^8)$ finite field. This gives AES a quite neat algebraic description.

However, they can also be seen as byte operations, and we'll look at it mainly as a byte operation, since we don't really want to get into math here. But if you really want to get deep into crypto, then you will certainly need to learn about finite fields [11]. They get more important in public key crypto, where we actually use *difficult* mathematical problems to protect our data.

The basic building blocks of the AES cipher are as follows:
Sub Bytes - A non-linear substitution, the AES S-boxes
Shift Rows - Shifts the rows of the AES state
Mix Column - Mixes columns of the AES state, making each result cell a combination of other cells
AddRoundKey - Mixes the input AES state with the current round key

As you can see, like in DES we have S-boxes; we have transpositions (Shift Rows), a mixing operation (Mix Column) and an operation to mix the data and the key. An AES encryption consists of the following steps:
Initial round:
AddRoundKey
R-1 rounds:
Sub Bytes
Shift Rows
Mix Columns
AddRoundKey
Final round (without Mix Columns):
Sub Bytes
Shift Rows
AddRoundKey

So, we have an initial AddRoundKey step, which mixes input data with the *0th* round key. Then, *R-1* (9, 11 or 13) identical rounds take place, and at the end a final round is applied.

**Security Contribution By RSA**

Clouds provide on-demand access to computing utilities, an abstraction of unlimited computing resources, and support for on-demand scale up, scale down and scale out. In this talk I will cover the "why", "what" and "how" of Cloud computing and explore the landscape of cloud services [12] [13]. Cloud computing allows consumers and businesses to use applications without installation and access their personal files at any computer with internet access. The RSA an algorithm is provides the high security in high potential data encryption methodology; the clouds are working to maintain the situation of the security and handover the position. Cloud computing is a comprehensive solution that delivers IT as a service. It is an Internet-based computing solution where shared resources are provided like electricity distributed on the electrical grid. Computers in the cloud are configured to work together and the various applications use the collective computing power as if they are running on a single system [14][15]. The flexibility of cloud computing is a function of the allocation of resources on demand. In future work, this technology allows for much more efficient computing by centralizing storage, memory, processing and bandwidth. RSA algorithm is very help full to increase the performance on cloud in cloud computing.

**AES Algorithm Prototype**

Strong encryption technology is a core technology for protecting data in transit to and from the cloud as well as data stored in the cloud. The main aim of encryption technology is to provide safety, security to data. In current era, Microsoft allows up to five security accounts per client and one can use these different accounts to create different zones. But available techniques are not enough to provide security to data in cloud [16] [17]. It is a demand and necessity of the data owners/providers/users to get high security for data. AES technique comes with 128/192/256 bits operations. In the previous section the general working of AES is explained. In this section we will see the proposed model of AES with cloud computing. Fig 1 shows the basic model [18] [19]. In this, data sender sends the data to AES model which applies the process to get the encrypted data which then sent to cloud service provider [20][21][22]. Cloud service provider will send the stored data which is in encrypted format to the requested site. At the receiver side AES model will decrypt the data.

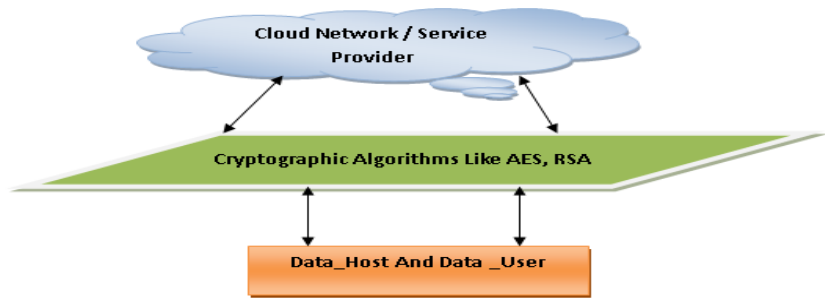

Figure 1 Cloud Model with Cryptography

## Conclusion

To improve the security for the data we can add biometrics features to the proposed algorithm. Now a day's depending upon the particular cloud provider, you can create multiple accounts with different keys. On Amazon Web Service, you can create multiple keys and rotate those keys during different sessions. In this paper we have proposed AES method for security of data in cloud computing. AES method can be most safe, trusted technique to provide security to data in clouds compared to other available security techniques in cloud computing. AES is used in data transit operations on Internet. Here we have suggested data storage, transmit, and receive operations using AES. It can improve data reliability, integrity with cloud-storage. Although encryption protects your data from unauthorized access, it does nothing to prevent data loss. Indeed, a common means for losing encrypted data is to lose the keys that provide access to the data. In this paper we have suggested AES and RSA technique for data security in clouds. Proposed algorithm is under implementation.